\def\year{2022}\relax
\documentclass[letterpaper]{article} 
\usepackage{aaai22}  
\usepackage{times}  
\usepackage{helvet}  
\usepackage{courier}  
\usepackage[hyphens]{url}  
\usepackage{graphicx} 
\usepackage{natbib}  
\usepackage{caption} 
\DeclareCaptionStyle{ruled}{labelfont=normalfont,labelsep=colon,strut=off} 
\usepackage{algorithm}
\usepackage{algorithmic}

\usepackage{newfloat}
\usepackage{listings}
\floatstyle{ruled}
\newfloat{listing}{tb}{lst}{}
\floatname{listing}{Listing}
\nocopyright
\title{Exploiting the Right: Inferring Ideological Alignment in Online Influence Campaigns Using Shared Images}
\author{
   Amogh Joshi and Cody Buntain\textsuperscript{\rm 1} \\
}

\affiliations{
   \textsuperscript{\rm 1}University of Maryland, College Park\\
   College Park, MD, 20742\\
   joshi.amoghn@gmail.com, cbuntain@umd.edu
}

\begin{document}

\maketitle

\begin{abstract}
This work advances investigations into the visual media shared by agents in disinformation campaigns by characterizing the images shared by accounts identified by Twitter as being part of such campaigns.
Using images shared by US politicians' Twitter accounts as a baseline and training set, we build models for inferring the ideological presentation of accounts using the images they share.
Results show that, while our models recover the expected bimodal ideological distribution of US politicians, we find that, on average, four separate influence campaigns -- attributed to Iran, Russia, China, and Venezuela -- all present conservative ideological presentations in the images they share.
Given that prior work has shown Twitter accounts used by Russian disinformation agents are ideologically diverse in the text and news they share, these image-oriented findings provide new insights into potential axes of coordination and suggest these accounts may not present consistent ideological positions across modalities.
\end{abstract}

\section{Introduction}

Court filings from the Department of Justice and similar sources allege that foreign interests, especially among Russian agents of disinformation, have sought to influence politics in the United States and abroad. 
These filings describe a sophisticated, multifaceted influence campaign, crossing platform boundaries and leveraging multiple modalities of content, from text to audio to visual.
While identifying such coordinated, malevolent, and covert campaigns remains a priority for the platforms and researchers studying platform integrity, much of the methods proposed for this task focus primarily on textual content and network interactions despite knowing these campaigns cover the spectrum of content modality.
Consequently, the manners in which these campaigns leverage content types that are difficult to characterize from text -- and images in particular -- remain largely understudied questions.

A key element of this understudied area is whether these agents of disinformation present a \emph{consistent} identity across these modalities: Since generating imagery is more expensive than generating text, such agents may appear more coordinated in the imagery they use compared to text, and the imagery they share may be inconsistent with the text they share -- in fact, more sophisticated campaigns might actually exploit this deviation and inconsistency as a method for pre-propaganda, as suggested by \cite{doi:10.1177/1940161220912682} in studying Russian disinformation agents' use of YouTube videos in the lead up to the 2016 US presidential election.

Before we can answer questions about multi-modal consistency, however, we first require a method for characterizing image-based sharing in political contexts.
This paper addresses this open question by developing multiple models for inferring political ideologies of accounts based on the \emph{images} they share rather than the text they write, domains they amplfy, or individuals with whom they interact (as those modalities have been studied at length, e.g., \cite{Temporao2018}, \cite{Eady2020}, \cite{Barbera2015}).
To train and validate these visual models for ideology, we rely on images shared on Twitter by over 600 US congresspeople, for whom we have well-accepted ideological positions using \texttt{DW-NOMINATE} scores taken from VoteView.org \cite{Lewis2021}.
We then apply these models to images shared by accounts in \emph{four known online influence campaigns}  -- identified and attributed to Iranian, Russian, Chinese, and Venezuelan states in Twitter's Election Integrity datasets -- and evaluate their ideological presentations.
This work examines these visually driven ideological presentations to answer the following research questions:

\begin{itemize}
    \item \textbf{RQ1} -- For images shared by US congresspeople in Twitter, how well do predictive models using these images correlate with a congressperson's ideological position?
    \item \textbf{RQ2} -- Does the distribution of ideologies reflect the bi-modal partisan divide in the US?
    \item \textbf{RQ3} -- When applied to influence campaign accounts, does one ideological position dominate the other?
\end{itemize}

Results suggest that, indeed, a US congressperson's Twitter account and its use of images does strongly correlate with that congressperson's NOMINATE-based ideological position, and these distributions are similarly bimodal, as we see in the NOMINATE dataset.
Applying these models to the four Twitter-based online influence campaigns shows all four campaigns present a slightly conservative bias in the images they share.
Contributions from this work chiefly include preliminary insights into ideological presentation laying a foundation for methods to evaluate account-level ideological presentation across multiple modalities.

\section{Methods}

At a high level, we answer these questions by using an off-the-shelf image embedding model to characterize visuals shared by a given account.
After averaging an account's image embeddings, we train supervised models to predict liberal/conservative ideological positions from these embeddings, using behavior of US congresspeople.
Applying these trained ideology inference models to known influence campaigns then allows us to infer the ideological position of accounts from these campaigns.

\subsection{Image Data, Politicians, and Influence Campaigns}

All image data for the four identified influence campaigns are sourced from Twitter's own election integrity datasets.\footnote{\url{https://transparency.twitter.com/en/reports/information-operations.html}}
This large archive provides textual data files containing the images shared by each account in the archive.
For a baseline of images shared in political contexts from authentic actors, we use a collection of images from US congresspeople made available from~\citet{first_paper}, which contains approximately 20 images shared by 705 US congresspersons' Twitter accounts.
Table \ref{tab:account_statistics} shows the number of accounts sharing images across these four influence campaigns and politicians' datasets.

\begin{table}[h]
    \scriptsize
    \centering
    \begin{tabular}{l r r r r r}
    \hline \\[-1em]
         & Iran & Russia & China & Venezuela & Politicians \\\hline\\[-1em]
Accounts & 435 & 1775 & 1783 & 963 & 705\\\hline
    \end{tabular}
    \caption{Accounts Per Group. Politicians represent US congresspeople, whereas the Iran, Russia, China, and Venezuela have been identified as disinformation agents by Twitter.}
    \label{tab:account_statistics}
\end{table}

%
%

\subsection{Determining Ideology of Influence Campaigns}

To assess political ideology of an influence campaign account from its shared imagery, we construct two regression models, both of which are trained against features from US congresspeople's images and their DW-NOMINATE ideology scores.
To featurize images, we use a pre-trained ResNet50 deep learning model to generate $2048$-dimensional embedding for each image.
In the first model (Model 1), we use an account's distribution of images across eight clusters of image types identified in~\citet{first_paper} and use a random forest regression model to predict an account's ideology from this distribution. 
This first model collapses the feature space of images into a small set of image types, and while~\citet{first_paper} shows a subset of these types correlate with ideological positions, this limited feature space may be too coarse.
As such, we include a second model (Model 2) that uses the raw image embeddings as features. 

\begin{figure}
    \centering
    \includegraphics[width=\columnwidth]{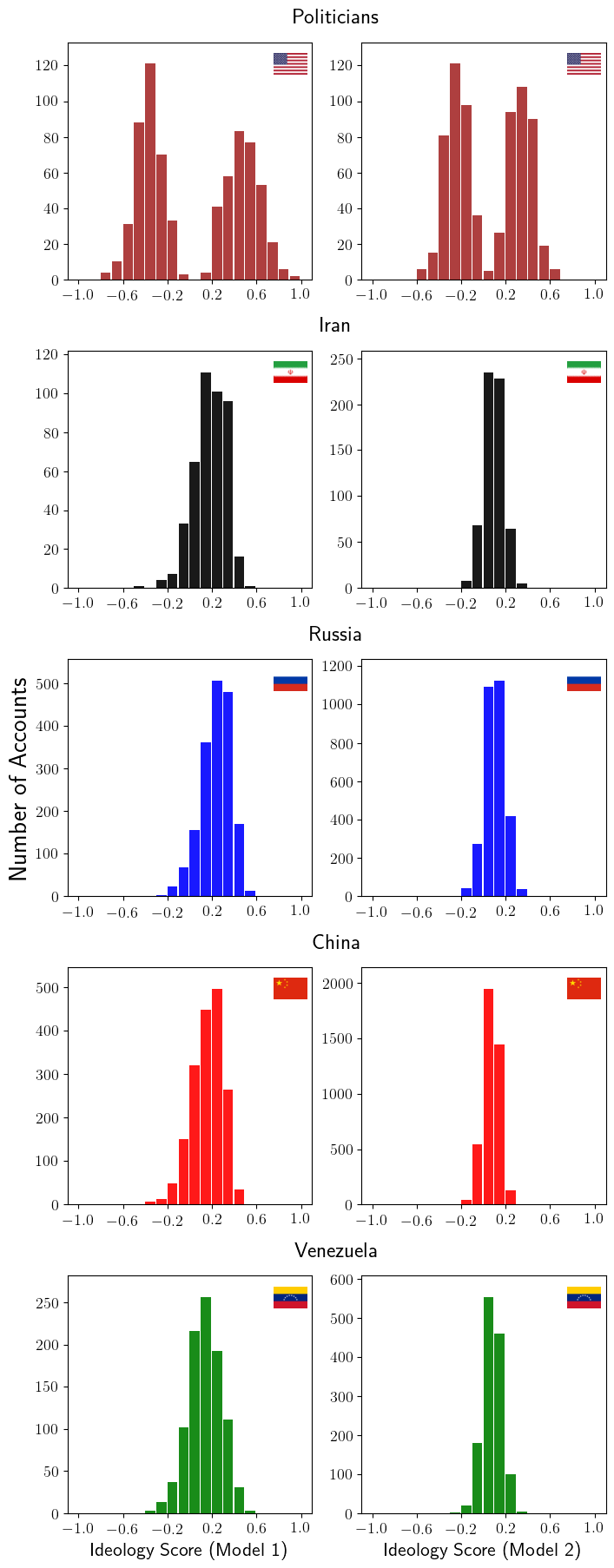}
    \caption{Ideology Distributions for Influence Campaign Accounts by Source. Model 1 is trained on cluster proportions, and Model 2 is trained on average image embeddings.}
    \label{fig:ideology_distribution}
\end{figure}

\section{Results}

%
%

\begin{figure*}[h]
    \centering
    \includegraphics[width=2.1\columnwidth]{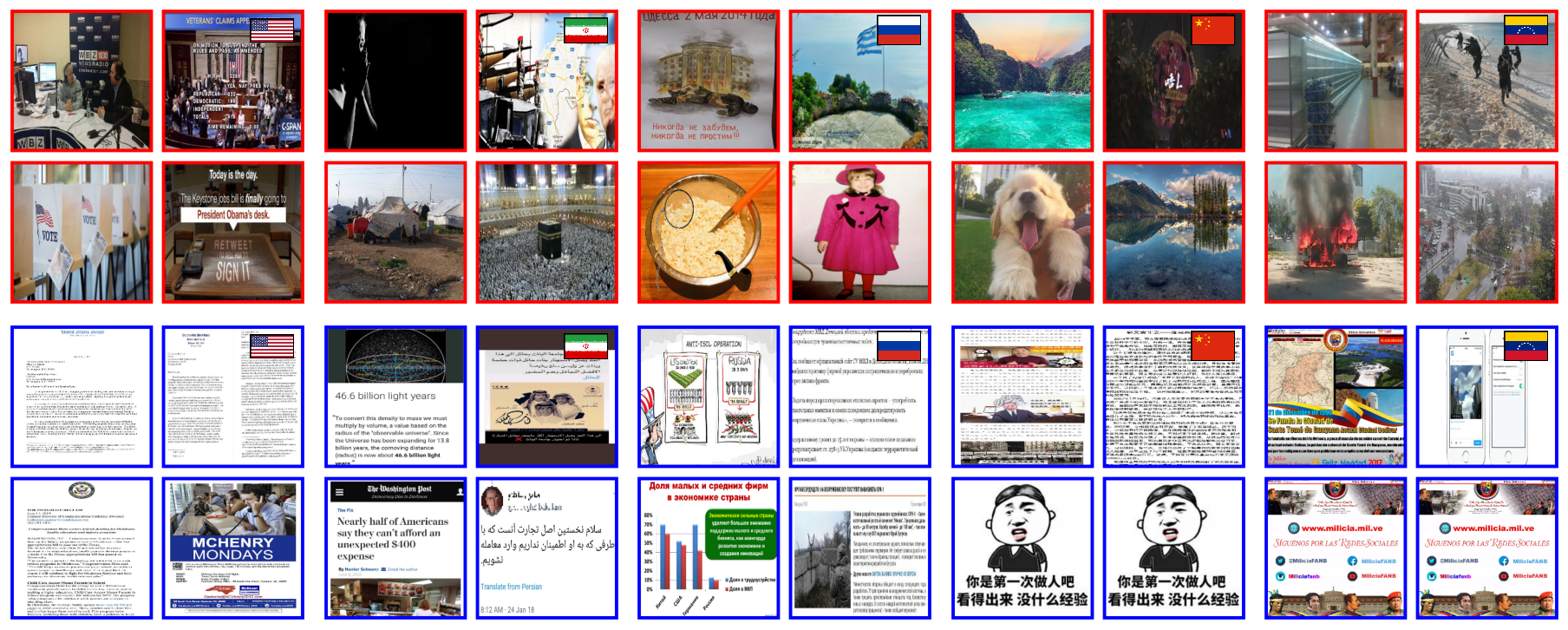}
    \caption{Sample Imagery from Conservative (Red Border) and Liberal (Blue Border) Clusters. The flag in the top right corners indicates the country from which the images are taken.}
    \label{fig:cluster_image_samples}
\end{figure*}

Following training on politicians' images and ideologies, our first regression model obtains an $R^2=0.869$, and Model 2 achieves $R^2=0.885$, suggesting an answer to \textbf{RQ1}: Image characterizations correlate strongly with ideological positions. 
Both models of these politicians' image-based ideologies produce the expected bimodal distributions, as shown in the top panel of Figure \ref{fig:ideology_distribution}, suggesting an affirmative answer for \textbf{RQ2}.

This answer to our second research question is particularly germane as we apply these inferential models to influence campaign accounts, as shown in the bottom four panels of Figure \ref{fig:ideology_distribution}.
All four campaigns appear to have a unimodal distribution largely centered around a moderately conservative ideology. 
Table~\ref{tab:cluster_image_distribution} presents mean ideological positions for all five groups across both models, showing that, on average, influence-campaign accounts present a more conservative position based on the images they share (\textbf{RQ3}).


\begin{table}[h]
    \scriptsize
    \centering
    \begin{tabular}{l l r r r r r}
    \hline\\[-1em]
       & & Iran & Russia & China & Venezuela & Politicians \\\hline\\[-1em]
\textbf{Model 1} & \textit{Mean} & 0.191 & 0.244 & 0.167 & 0.140 & 0.051 \\
& \textit{Std. Dev.} & 0.145 & 0.130 & 0.138 & 0.147 & 0.313 \\\hline\\[-1em]
\textbf{Model 2} & \textit{Mean} & 0.096 & 0.108 & 0.075 & 0.085 & 0.052 \\
& \textit{Std. Dev.} & 0.087 & 0.089 & 0.071 & 0.083 & 0.311\\\hline
    \end{tabular}
    \caption{Average Ideology for Each Influence Campaign. Both models illustrate an average conservative lean across all four campaigns, with the Russian campaign presenting the most conservative average account.}
    \label{tab:cluster_image_distribution}
\end{table}

We also inspect a sample of images taken from liberal and conservative clusters identified in~\citet{first_paper}, as shown in Figure~\ref{fig:cluster_image_samples}.
Conservative-leaning imagery shared by influence campaigns appears to follow those used by American politicians as discussed in~\citet{first_paper}: politically-charged images of demonstrations, landscapes, American patriotism, and military actions. 
Liberal-leaning imagery consists largely of document-based visuals, screenshots, and assorted infographics.

\section{Conclusion}

These preliminary results demonstrate the ability to predict the political ideology of US politicians and targeted influence campaigns alike. 
We observe that most political influence campaign accounts present a moderately conservative slant, with few accounts sharing liberally oriented visuals. 
Given that prior work has shown Twitter accounts used by Russian disinformation agents are ideologically diverse in the text and news they share, these image-oriented findings are surprising and potentially provide new insights into axes of coordination and suggest these accounts may not present consistent ideological positions across modalities.

\bibliography{aaai22}

\end{document}